\documentclass[twocolumn,aps,prl,preprintnumbers]{revtex4}
\usepackage{bbm}
\usepackage[latin9]{inputenc}
\usepackage{amsmath}
\usepackage{amssymb}
\usepackage{graphicx}
\usepackage{mathrsfs}
\usepackage{amsfonts}
\usepackage{amsthm}
\usepackage{color}
\usepackage{txfonts}
\usepackage[colorlinks=true,citecolor=blue,linkcolor=blue,urlcolor=blue,anchorcolor=blue]{hyperref}%
\usepackage{pifont}
\hypersetup{colorlinks=true,citecolor=blue,linkcolor=blue,urlcolor=blue}
\providecommand{\U}[1]{\protect\rule{.1in}{.1in}}
\makeatletter
\@ifundefined{textcolor}{}
{
\definecolor{BLACK}{gray}{0}
\definecolor{WHITE}{gray}{1}
\definecolor{RED}{rgb}{1,0,0}
\definecolor{GREEN}{rgb}{0,1,0}
\definecolor{BLUE}{rgb}{0,0,1}
\definecolor{CYAN}{cmyk}{1,0,0,0}
\definecolor{MAGENTA}{cmyk}{0,1,0,0}
\definecolor{YELLOW}{cmyk}{0,0,1,0}
}

\makeatother
\begin{document}
\title{Cavity-Induced Strong Magnon-Magnon Coupling in Altermagnets}
\author{Zhejunyu Jin}
\author{Huanhuan Yang}
\author{Zhaozhuo Zeng}
\author{Yunshan Cao}
\author{Peng Yan}
\email[Corresponding author: ]{yan@uestc.edu.cn}
\affiliation{School of Physics and State Key Laboratory of Electronic Thin Films and Integrated Devices, University of
Electronic Science and Technology of China, Chengdu 610054, China}

\begin{abstract}
Long-distance strong coupling between short-wavelength magnons remains an outstanding challenge in quantum magnonics, an emerging interdiscipline between magnonics and quantum information science. Recently, altermagnets are identified as the third elementary class of magnets that break the time-reversal symmetry without magnetization and thus combine characteristics of conventional collinear ferromagnets and antiferromagnets. In this work, we show that cavity photons can mediate the long-distance strong coupling of exchange magnons with opposite chiralities in altermagnets, manifesting as an anticrossing of the magnon-polariton spectrum in the extremely dispersive regime. The predicted effective magnon-magnon coupling strongly depends on the magnon propagation direction, and is thus highly anisotropic. Our findings are intimately connected to the intrinsic nature of altermagnetic magnons, i.e., chirality-splitting-induced crossing of exchange magnons, which has no counterpart in conventional ferromagnets or antiferromagnets, and may open a new path way for magnon-based quantum information processing in altermagnets.
\end{abstract}

\maketitle

\textit{Introduction.---}Magnons (quanta of spin waves) have been extensively explored for wave-based sensing and computing concepts, due to their long lifetime and high tunability \cite{Kruglyak2010,Chumak2015,Pirro2021}. The compatibility between magnons and diverse quantum platforms such as qubit \cite{Tabuchi2015,Mq2022,Mq2023}, phonon \cite{Agrawal2013,Streib2019,Bozhko2020}, and photon \cite{Bai2011,Braggio2016,Harder2018} further amplifies the advantages of utilizing magnons as a carrier for quantum information processing, constituting quantum magnonics \cite{Yuan2022}. Coherent transfer of magnetic information between two magnonic systems demands a strong magnon-magnon coupling, which is usually generated by dipolar interaction \cite{Shiota2020}, interlayer exchange \cite{Chen2018,Ndiaye2017,Sklenar2021}, and in-plane anisotropy \cite{Liensberger2019}, etc. However, limited by the effective range of these magnetic interactions, they can only mediate the short-distance coupling between magnons. It is noted that the coherent magnon-photon interaction has been reported in hybrid cavity-magnet systems \cite{Soykal2010,Yuan2017}. Cavity photons can also couple to two magnons simultaneously over a long distance, inducing a nonlocal magnon-magnon interaction in both ferromagnets and antiferromagnets \cite{Lambert2016,Rameshti2018,Grigoryan2019,Nair2022, Zhang2023,Johansen2018}. Nevertheless, past studies focused on the long-wavelength limits, i.e., magnetostatic magnons. The nonlocal strong coupling between short-wavelength magnons remains an outstanding challenge in the community.

\begin{figure}[htbp]
  \centering
  % Requires \usepackage{graphicx}
  \includegraphics[width=0.48\textwidth]{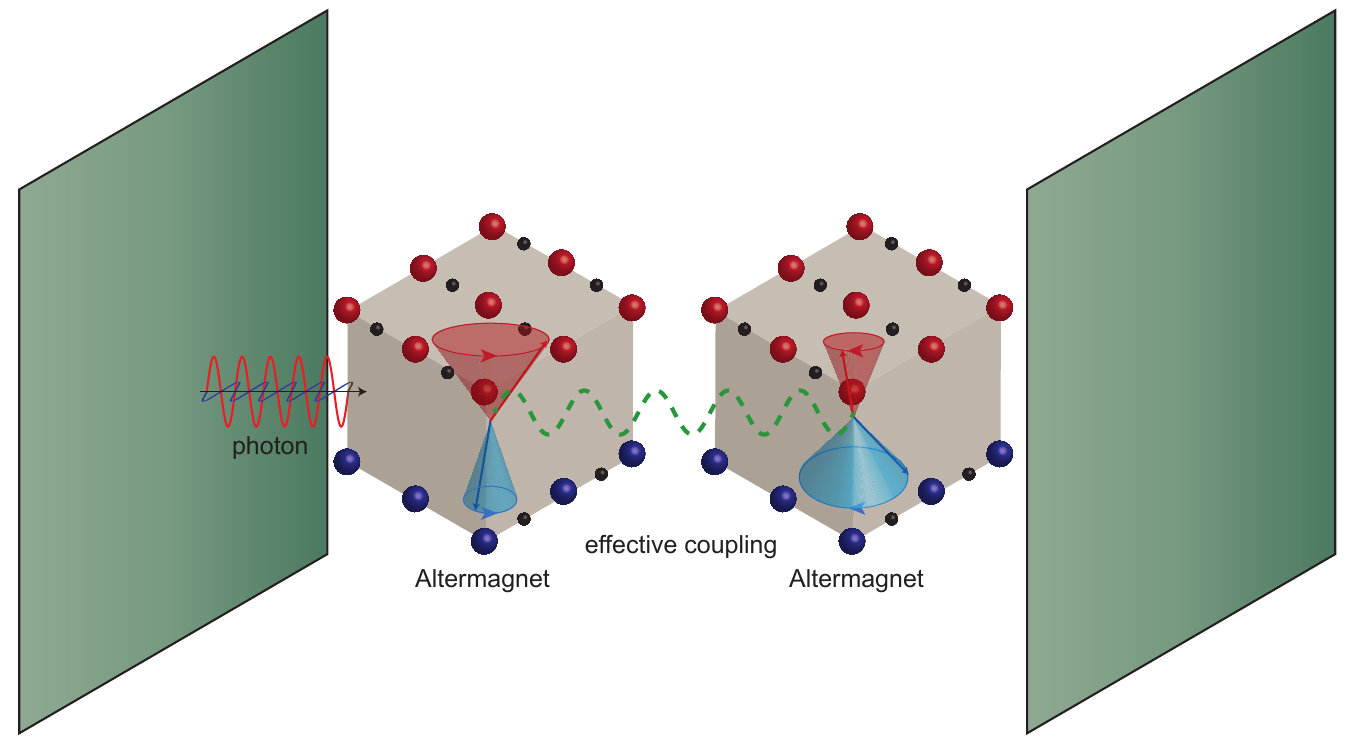}\\
 \caption{Schematics of the cavity-induced long-distance magnon-magnon coupling in altermagnets. The effective coupling (dashed green wavy line) between two magnons with opposite chiralities is mediated by the cavity photon.}\label{fig1}
\end{figure}

Recently, an emerging class of magnets dubbed altermagnet was identified, which maintains zero net macroscopic magnetization but exhibits a surprisingly large spin-splitting \cite{Smejkal1,Smejkal2,Smejkal3,Mazin2023,Turek2022,Feng2022,Ouassou2023,SZhang2023,Zhou2023,Hariki2023,Sun2023,Bai2023}. These peculiar features are protected by the combined spin and lattice symmetries, i.e., the spin-space inversion and crystallographic-space rotation \cite{Smejkal1,Smejkal2}, which make altermagnets a promising platform for quantum magnonics. It has been shown that the chiral-splitting effect induces the crossing of infinitely-long-wavelength magnons (${\bf k}=0$, where ${\bf k}$ is the wavevector) with opposite chiralities \cite{Smejkal2}. But it brings about two open issues for realizing strong nonlocal coupling between exchange magnons: (i) How to push the crossing point to the short-wavelength region; and (ii) How to realize the coupling between these two orthogonal magnon modes, which was forbidden by angular momentum conservation, a consequence of rotational invariance or isotropy.

In this Letter, we demonstrate the cavity-induced long-distance strong coupling of exchange magnons in altermagnets (Fig. \ref{fig1}). We show that a perpendicular magnetic field can lead to an upward (downward) shift of the magnon branch with a low (high) group-velocity, resulting in an unavoided level crossing at a finite wave number (${\bf k}\neq 0$). Then, we find that the magnon degeneracy can be lifted by placing the altermagnet in a photonic cavity, manifesting as an anticrossing in the magnon-polariton spectrum. Surprisingly, it is observed that the photon frequency is orders of magnitude higher than that of magnons at the anticrossing point, indicating a highly dispersive magnon-photon coupling. By utilizing a second-order perturbation theory, we derive the analytical formula of the cavity-induced effective magnon-magnon coupling. We show that the indirect coupling strongly depends on the magnon propagation direction due to the anisotropic nature of the exchange interaction in altermagnets. Our results open the door for exploring quantum magnonics based on the altermagnetism.

\begin{figure}[htbp]
  \centering
  % Requires \usepackage{graphicx}
  \includegraphics[width=0.48\textwidth]{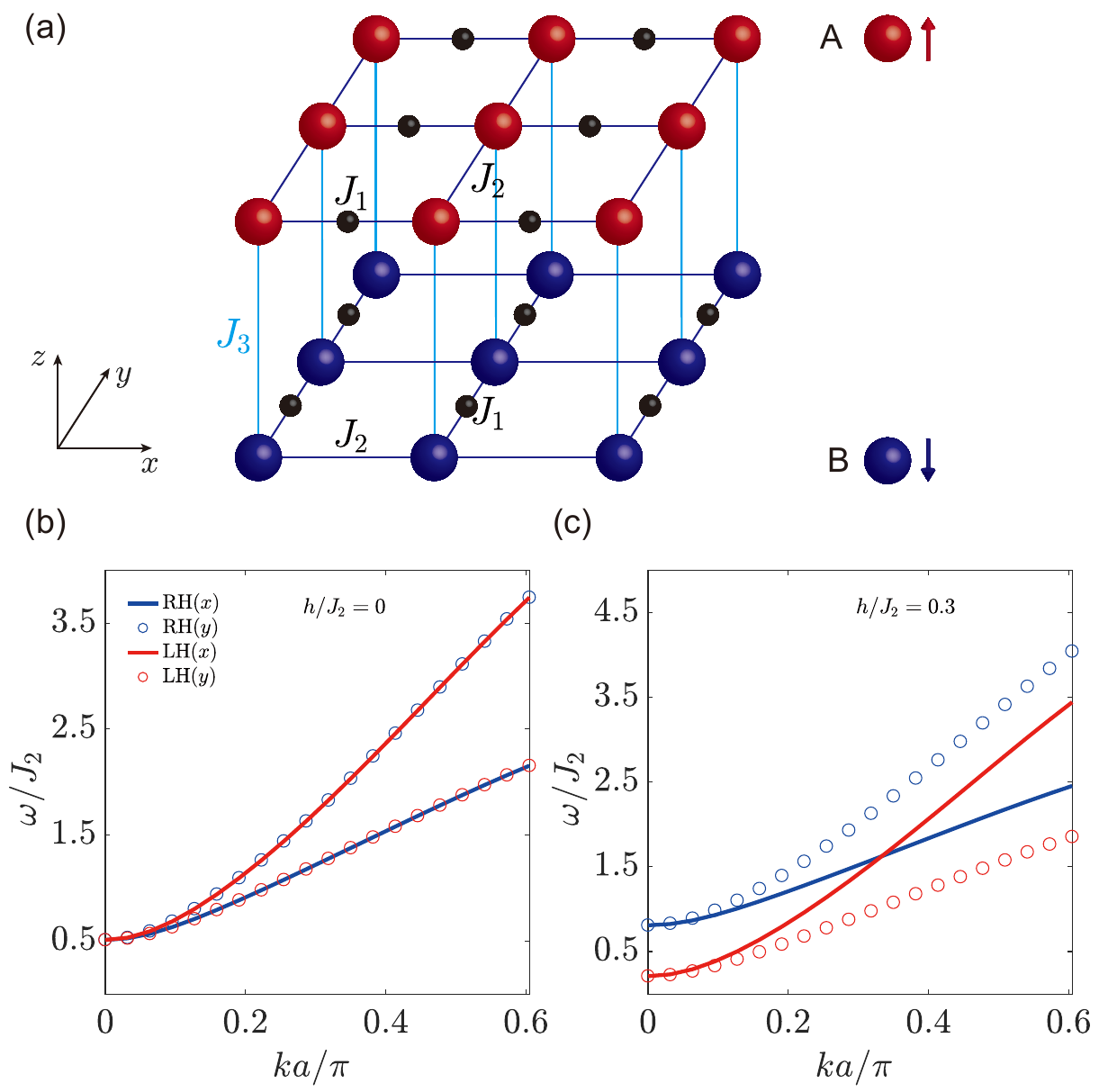}\\
 \caption{(a) Schematic illustration of a two-sublattice altermagnet. Spin-up and-down atoms are labeled by red and blue spheres, respectively. The black spheres correspond to nonmagnetic atoms. Magnon dispersion for $h/J_2=0$ (b) and $h/J_2 =0.3$ (c). Curves and circles represent the magnons propagating along $x$ and $y$ direction, respectively. The handedness of magnons is distinguished by contrast color (blue and red). Labels $x$ and $y$ represent the propagation direction of magnons.}\label{fig2}
\end{figure}

\textit{Chiral splitting of magnons in altermagnets.---}Let us consider a two sublattice altermagnet with anisotropic intralayer exchange interactions. The spin Hamiltonian reads
\begin{equation}\label{Eq1}
\begin{aligned}
{\mathcal H}_{\text{alter}}=&-S^2\sum_{i,j}\big[J_{1}{\bf s}_{i,j}^A\cdot{\bf s}_{i+1,j}^A+J_{2}{\bf s}_{i,j}^B\cdot{\bf s}_{i+1,j}^B\\
&+J_{2}{\bf s}_{i,j}^A\cdot{\bf s}_{i,j+1}^A+J_{1}{\bf s}_{i,j}^B\cdot{\bf s}_{i,j+1}^B+J_3{\bf s}_{i,j}^A\cdot{\bf s}_{i,j}^B\\
&+S^{-1}{\bf h}\cdot({\bf s}_{i,j}^A+{\bf s}_{i,j}^B)+K({\bf s}_{i,j}^A\cdot{\bf z})^2+K({\bf s}_{i,j}^B\cdot{\bf z})^2\big],\\
\end{aligned}
\end{equation}
where $J_{1,2}>0$  represents the intralayer ferromagnetic exchange coupling strength, $J_3<0$ is the interlayer antiferromagnetic exchange coupling coefficient, ${\bf h}$ and $K$ denote the external magnetic field and the magnetic anisotropy constant, respectively, ${\bf s}_{i,j}^A$ and ${\bf s}_{i,j}^B$ are the normalized spins on sites $(i,j)$ of sublattices $A$ and $B$, respectively, and $S$ is the length of the spin vector. Figure \ref{fig2}(a) shows the crystal structure of a two-sublattice altermagnet. Under a combined operation of two-fold spin-space rotation $\mathcal{C}_{s,2}$ (${\bf s}^A\rightarrow {\bf s}^B$) and four-fold crystallographic-space rotation $\mathcal{C}_4$ $(i\rightarrow j$ and $J_1\rightarrow J_2)$, we find that Hamiltonian \eqref{Eq1} is invariant and thus respects the symmetry of altermagnet \cite{Smejkal1}. We then obtain the dispersion relation of altermagnetic magnons \cite{SM}
\begin{equation}\label{Eq4}
\begin{aligned}
\omega_{{\bf k},\pm}&=\pm c_1({\bf k})+c_2({\bf k}),
\end{aligned}
\end{equation}with $c_1({\bf k})=S^2\Big \{h/S+(J_1-J_2)\big[\cos(k_ya)-\cos(k_xa)\big]\Big \}$ and $c_2({\bf k})=S^2\Big \{2K-(J_2+J_1)\big[\cos(k_xa)+\cos(k_ya)-2\big]\Big \}^{1/2}\Big \{2K-2J_3-(J_1+J_2)\big[\cos(k_xa)+\cos(k_ya)-2\big]\Big \}^{1/2}$. Here, $a$ is the lattice constant and $\pm$ corresponds to right-handed (RH) and left-handed (LH) magnon modes, with respect to the $z$ axis, respectively. In what follows, we use the following parameters to calculate the spectrum: $J_1= 0.4J_2$, $J_3 = -1.25J_2$, $K = 0.05J_2$, and $S=1$, if not stated otherwise. Different from the antiferromagnet, the degeneracy of RH and LH magnons is broken even in the absence of the external magnetic field, resulting in the unequal magnon group velocity, as shown in Fig. \ref{fig2}(b). In addition, it is noted that the spectrum of RH magnons propagating along the $x$ axis is identical to LH magnons propagating along the $y$ axis, and vice versa. When a perpendicular magnetic field is applied, the energy degeneracy at $k=0$ is removed ($k=|{\bf k}|$). However, we observe that, for magnons propagating along $x$-direction, the branch with a lower group-velocity shifts upward, while the one with a higher group-velocity shifts downward. This results in an unavoided level crossing at a finite wave number, i.e., $k\neq 0$ [see red and blue curves in Fig. \ref{fig2}(c)]. Such feature does not exist for magnons propagating along the $y$ direction, which is an indication of the crossing anisotropy and will be discussed below. A level crossing usually means the absence of coupling. Particularly for two magnons with opposite handedness, their direct coupling was thought to be forbidden by angular momentum conservation, a consequence of rotational invariance.  As to the long-distance coupling between exchange magnons, it even looks like a dilemma at first sight because the exchange interaction is extremely short-ranged. Next, we tackle this problem in cavities.

\textit{Cavity-induced magnon-magnon coupling.---}To this end, we consider the altermagnet embedded in a photonic cavity, described by following Hamiltonian
\begin{equation}\label{Eq5}
\begin{aligned}
\mathcal{H}&={\mathcal H}_{\text{alter}}+{\mathcal H}_{\text{ph}}+{\mathcal H}_{\text{int}},\\
{\mathcal H}_{\text{ph}}&=\frac{1}{2}\int(\epsilon_0{\bf E}^2+\frac{{\bf B}^2}{\mu_0})d{\bf r},\\
{\mathcal H}_{\text{int}}&=-\frac{1}{\mu_0}\sum_{i,j}({\bf s}_{i,j}^A+{\bf s}_{i,j}^B)\cdot
{\bf B},
\end{aligned}
\end{equation}
where $\mathcal{H}_{\text{ph}}$ is the photon Hamiltonian, $\mathcal{H}_{\text{int}}$ is the magnon-photon coupling, ${\bf E}$ and ${\bf B}$ are the electric and magnetic components of the electromagnetic wave, respectively, and $\epsilon_0$ and $\mu_0$ are vacuum permittivity and susceptibility, respectively. By using the Holstein-Primakoff transformation \cite{HP}: $s^{A,+}=\sqrt{2}a,s^{B,+}=\sqrt{2}b^\dag, s^{A,-}=\sqrt{2}a^\dag,s^{B,-}=\sqrt{2}b, s^A_z=1-a^\dag a,$ and $s^B_z=b^\dag b-1$,
\begin{figure}[htbp]
  \centering
  % Requires \usepackage{graphicx}
  \includegraphics[width=0.48\textwidth]{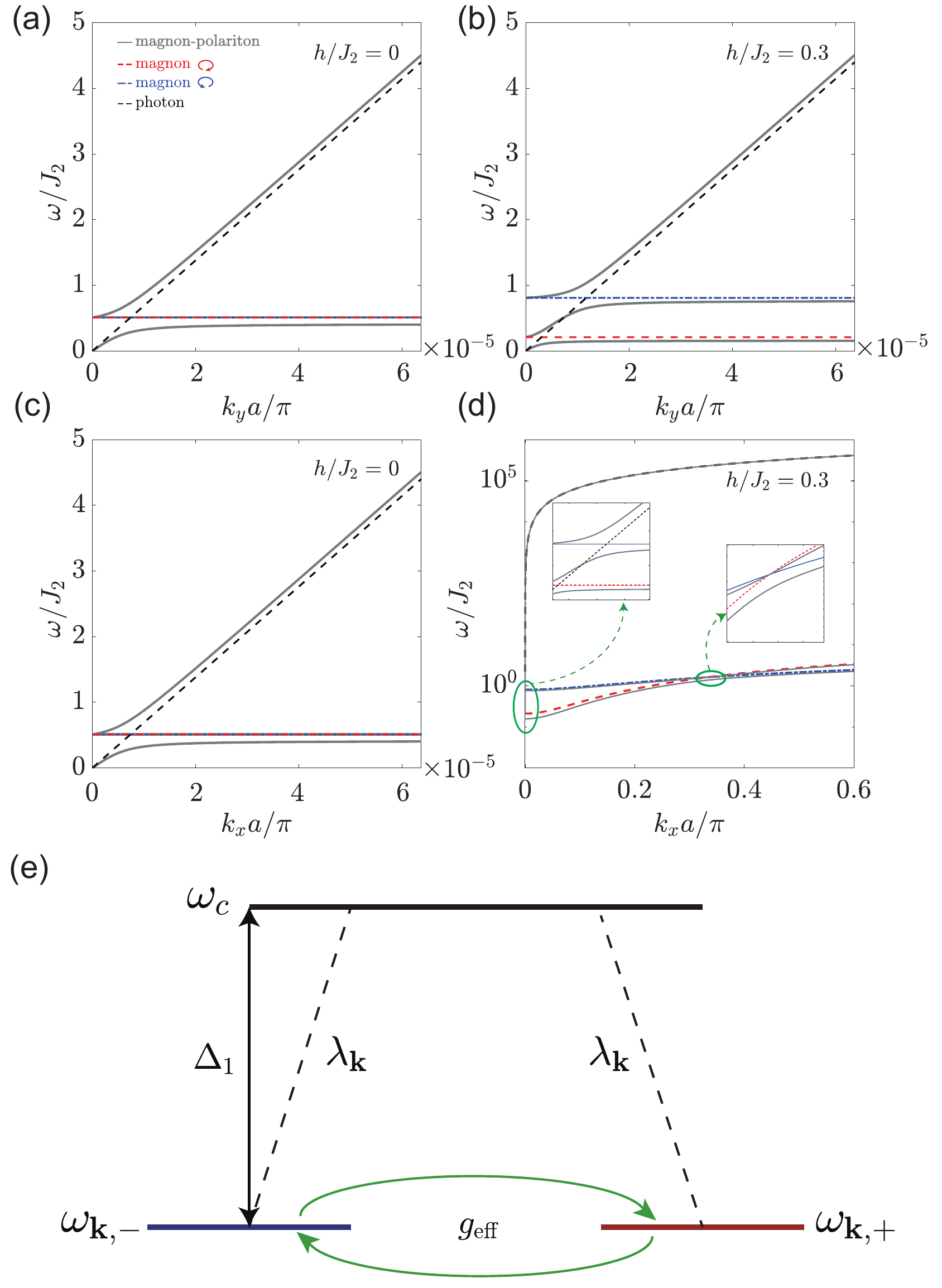}\\
 \caption{Dispersion relations of the magnon-polariton propagating along $y$-axis (a,b) and $x$-axis (c,d) with zero (a,c) and finite magnetic field (b,d). The dashed red, blue, and black lines represent uncoupled magnon modes and cavity mode, while the silver-gray lines represent the magnon-polariton. Insets show the details of the anticrossing in the spectrum. (e) Diagram of energy level and interaction scheme for two magnons with frequencies $\omega_{{\bf k},\pm}$ coupled to a cavity photon with frequency $\omega_c$, by exchanging virtual photons.}\label{fig3}
\end{figure}
where the spin operators $s^{m,\pm}=s^m_x\pm is^m_y$ with $m=A,B$, and $a(b)$ and $a^\dag(b^\dag)$ are the magnon annihilation and creation operators for sublattice $A(B)$, respectively, we can express the magnon Hamiltonian in the momentum space as
\begin{equation}\label{Eq7}
\begin{aligned}
\mathcal{H}_{\text{alter}}=&\sum_{{\bf k}}\Big \{\big [2J_{1}\cos(k_xa)+2J_{2}\cos(k_ya)+J_3-h\\
&-2J_{2}-2J_{1}+2K\big]a_{\bf k}^\dag a_{\bf k}+\big[2J_{2}\cos(k_xa)\\
&+2J_{1}\cos(k_ya)
+J_3+h-2J_{2}-2J_1\\
&+2K\big]b_{\bf k}^\dag b_{\bf k}
+J_3(a_{\bf k}b_{\bf k}+a_{\bf k}^{\dag}b_{\bf k}^\dag)\Big \}.
\end{aligned}
\end{equation}Here, we have assumed circularly-polarized photons, with the resulting photon Hamiltonian and magnon-photon coupling as follows
\begin{equation}\label{Eq8}
\begin{aligned}
\mathcal{H}_{\text{ph}}&=\sum_{{\bf k}}\omega_c(c_{\bf k}^\dag c_{\bf k}+\frac{1}{2}),\\
\mathcal{H}_{\text{int}}&=\sum_{{\bf k}}g_c(c_{\bf k}a_{\bf k}+c_{\bf k}^\dag a_{\bf k}^\dag+c_{\bf k}b_{\bf k}^\dag+c_{\bf k}^\dag b_{\bf k}),\\
\end{aligned}
\end{equation}
where $\omega_c=v_c|{\bf k}|$ is the photon dispersion relation with $v_c$ the speed of light and $g_c=\sqrt{\omega_c\mu_0 N/2 V}$ represents the magnon-photon coupling with the number of spins $N$ and the volume of cavity $V$ \cite{Yuan2017}. By utilizing the Bogoliubov transformation
\begin{equation}\label{Eq9}
\begin{aligned}
a_{\bf k}&=u_{\bf k}\alpha_{\bf k}+v_{\bf k}\beta_{\bf k}^\dag,\\
b_{\bf k}&=u_{\bf k}\beta_{\bf k}+v_{\bf k}\alpha_{\bf k}^\dag,\\
\end{aligned}
\end{equation}
where $u_{\bf k}=\sqrt{(\Delta_{\bf k}+1)/2}$, $v_{\bf k}=-\sqrt{(\Delta_{\bf k}-1)/2}$, and
\begin{equation}\label{Deltak}
\Delta_{\bf k}=\sqrt{\frac{1}{1-\Big\{\frac{J_3}{(J_1+J_2)[\cos(k_xa)+\cos(k_ya)]+J_3-2(J_1+J_2)-2K}\Big\}^2}},
\end{equation}the magnon Hamiltonian \eqref{Eq7} can be diagonalized  as
\begin{equation}\label{Eq10}
\begin{aligned}
\mathcal{H}_{\text{alter}}&=\sum_{{\bf k}}(\omega_{{\bf k},-}\alpha_{\bf k}^\dag \alpha_{\bf k}+\omega_{{\bf k},+}\beta_{\bf k}^\dag \beta_{\bf k}).
\end{aligned}
\end{equation}
The total Hamiltonian then can be recast as $\mathcal{H}=\psi^\dag \mathcal{M} \psi$ with the vector $\psi=(\alpha_{\bf k},\beta_{\bf k}^\dag, c_{\bf k}^\dag)^\dag$ and matrix
\begin{equation}\label{Eq11}
\mathcal{M}=\left(
    \begin{array}{ccc}
      \omega_{{\bf k},-} & 0 & \frac{\lambda_{\bf k}}{2} \\
      0 & \omega_{{\bf k},+} & \frac{\lambda_{\bf k}}{2} \\
      \frac{\lambda_{\bf k}}{2} & \frac{\lambda_{\bf k}}{2} & \omega_c \\
    \end{array}
  \right),
\end{equation}
with $\lambda_{\bf k}=2g_c(u_{\bf k}+v_{\bf k})$, leading to the following secular equation
\begin{equation}\label{Eq12}
\begin{aligned}
&4\omega^3-4(\omega_{{\bf k},+}+\omega_{{\bf k},-}+\omega_c)\omega^2+\lambda_{\bf k}^2(\omega_{{\bf k},+}
+\omega_{{\bf k},-})-4\omega_{{\bf k},+}\omega_{{\bf k},-}\omega_c\\
&+2\big [-\lambda_{\bf k}^2+2\omega_{{\bf k},+}\omega_{{\bf k},-}+2(\omega_{{\bf k},+}+\omega_{{\bf k},-})\omega_c\big ]\omega=0.
\end{aligned}
\end{equation}

By numerical solving Eq. \eqref{Eq12}, one obtains the dispersion relation of the coupled cavity-altermagnet. For the case without the magnetic field, near ${\bf k}=0$, one of the degenerated magnon bands is a dark mode that does not interact with the cavity photon, while the other one does, see Figs. \ref{fig3}(a) and (c). When the external field is present, the double degeneracy of magnons is removed, and both magnon modes couple with
the cavity photon, see Figs. \ref{fig3}(b) and (d). These features are close to the case in antiferromagnets due to the similar magnon dispersion in the long-wavelength limit \cite{Yuan2017,Yuan2020}. Strikingly unlike the antiferromagnet, we find an anticrossing gap emerging at $k_x=a^{-1}\arccos{\big [1-h/(J_2-J_1)\big ]}$ for magnons propagating along $x$ direction. This gap represents the effective magnon-magnon coupling. However, at this point, the photon frequency is much higher than magnon ($\omega_c/\omega_{{\bf k}, \pm}>10^5$) [see Fig. \ref{fig3}(d)]. The magnon-photon coupling is thus highly dispersive, where a resonant energy exchange between magnon and photon is not allowed. But magnons can exchange energy through virtual photons with the cavity, resulting in the indirect magnon-magnon coupling, as illustrated in Fig. \ref{fig3}(e). To obtain an analytical understanding, we derive the effective magnon-magnon coupling by using the following unitary transformation
\begin{figure}[htbp]
  \centering
  % Requires \usepackage{graphicx}
  \includegraphics[width=0.48\textwidth]{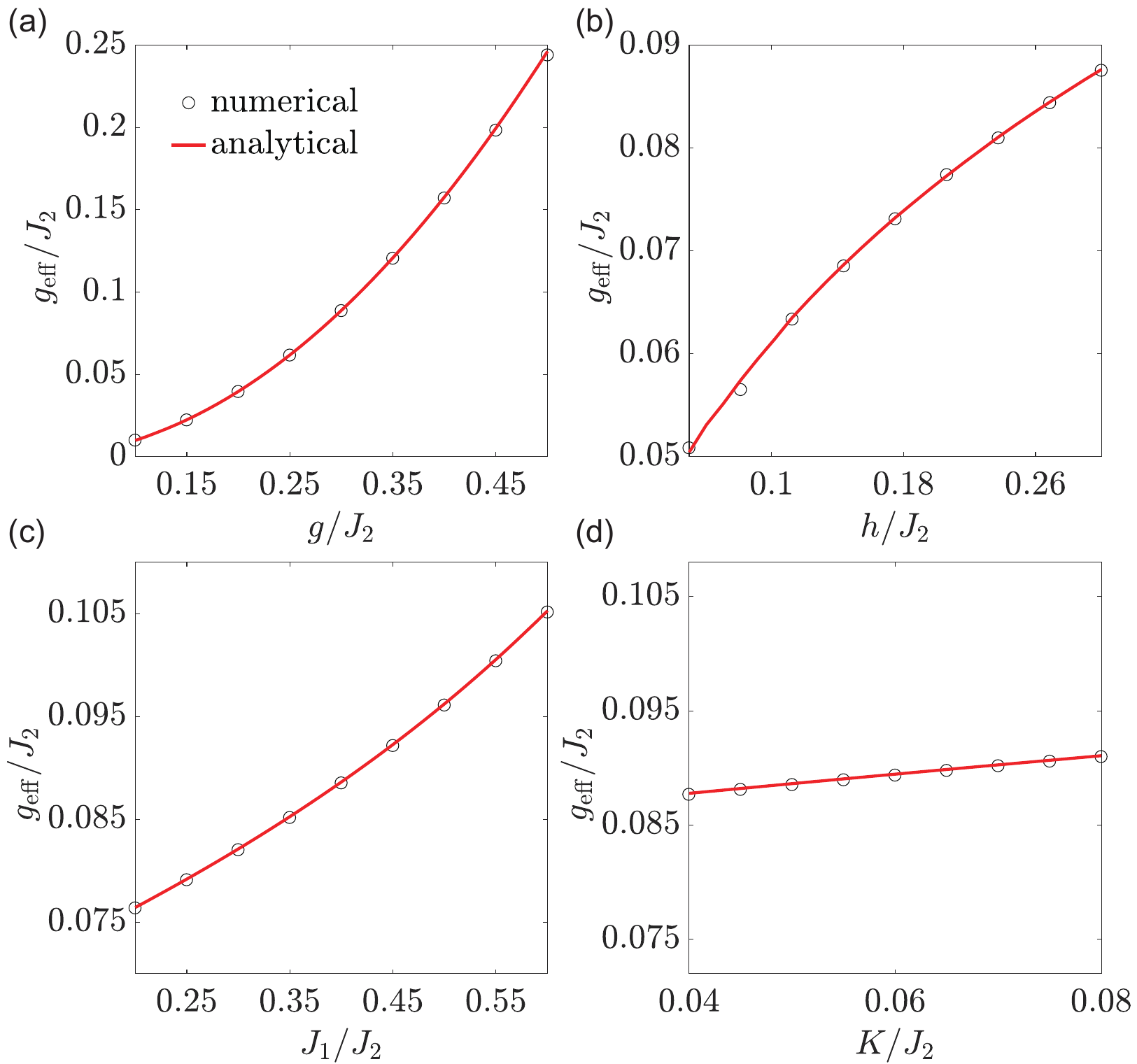}\\
 \caption{Analytical formula (red curves) and numerical results (black circles) of the effective magnon-magnon coupling $g_{\rm{eff}}$ as a function of the magnon-photon coupling strength $g=g_c/\sqrt{\omega_c/J_2}$ (a), magnetic filed $h$ (b), exchange ratio $J_1/J_2$ (c), and anisotropy constant $K$ (d).}\label{fig4}
\end{figure}
\begin{equation}\label{Eq13}
\begin{aligned}
U=\exp \Big[&\frac{\lambda_{\bf k} }{{2{\Delta_1}}}(c_{\bf k}^\dag {\beta_{\bf k}} - \beta_{\bf k}^\dag {c_{\bf k}} + c_{\bf k}^\dag {\alpha_{\bf k}} - \alpha_{\bf k}^\dag {c_{\bf k}})\\ &+\frac{\lambda_{\bf k}}{{2{\Delta _2}}}(c_{\bf k}^\dag \alpha_{\bf k}^\dag - {c_{\bf k}}{\alpha_{\bf k}} + c_{\bf k}^\dag \beta_{\bf k}^\dag  - {c_{\bf k}}{\beta_{\bf k}})\Big],
\end{aligned}
\end{equation}and expanding the transformed Hamiltonian to the second order of $\lambda_{\bf k}$
\begin{equation}\label{Eq14}
\begin{aligned}
\mathcal{H}_{\rm{eff}}&=U^\dag \mathcal{H}U=\Big[(\omega ' + \frac{{{\lambda_{\bf k}^2}}}{{4{\Delta _2}}} + \frac{{{\lambda_{\bf k}^2}}}{{4{\Delta _1}}})\alpha _{\bf k}^\dag {\alpha _{\bf k}} + (\omega ' + \frac{{3{\lambda_{\bf k}^2}}}{{4{\Delta _1}}}\\
  &- \frac{{{\lambda_{\bf k}^2}}}{{4{\Delta _2}}})\beta _{\bf k}^\dag {\beta _{\bf k}}+ ({\omega _c} - \frac{{{\lambda_{\bf k}^2}}}{{{\Delta _1}}})c_{\bf k}^\dag {c_{\bf k}}\Big] + g_{\rm{eff}}({\alpha _{\bf k}}\beta _{\bf k}^\dag  + {\beta _{\bf k}}\alpha _{\bf k}^\dag ),
\end{aligned}
\end{equation}
 where $\Delta_1=\omega_c-\omega'$ and $\Delta_2=-\omega_c-\omega'$ with the magnon frequency $\omega'$ at the crossing point. Here, $g_{\rm{eff}}=\frac{{{\lambda_{\bf k}^2}}}{{2{\Delta _1}}}$ describes the effective magnon-magnon coupling mediated by virtual photons. In the dispersive limit, a careful examination reveals that one of the hybridization modes passes through the degeneracy point where $\omega_+ = \omega_-=\omega'$, in sharp contrast to the observations in resonantly coupled systems \cite{Xu2019,Li2022}. This feature can also be seen from the three eigenvectors $\alpha_{\bf k}^\dag-\beta_{\bf k}, c_{\bf k}-\frac{\sqrt{2}}{2}(\alpha_{\bf k}^\dag+\beta_{\bf k})$, and $c_{\bf k}+\frac{\sqrt{2}}{2}(\alpha_{\bf k}^\dag+\beta_{\bf k})$, in which the state $\alpha_{\bf k}^\dag-\beta_{\bf k}$ is a magnon dark mode (also known as a subradiant mode) that is decoupled from the cavity \cite{Zhang2015}.

Figure \ref{fig4}(a) demonstrates that the effective coupling induced by virtual photon exchange is dominated by the second-order of the dispersive coupling $\lambda_{\bf k}$. Since $k_x$ at the anticrossing point is a nonlinear function of both the magnetic field $h$ and exchange ratio $J_1/J_2$, it indicates a nonlinear dependence of the effective coupling strength $g_{\rm{eff}}$ on $h$ and $J_1/J_2$, as shown in Figs. \ref{fig4}(b) and \ref{fig4}(c), respectively. The anisotropy constant $K$ merely modifies the magnon frequency, and hardly affects the coupling strength $g_{\rm{eff}}$ due to the huge frequency mismatch between magnons and cavity photons, as shown in Fig. \ref{fig4}(d). Our theory well explains the numerical calculations.

 \begin{figure}[htbp]
  \centering
  % Requires \usepackage{graphicx}
  \includegraphics[width=0.48\textwidth]{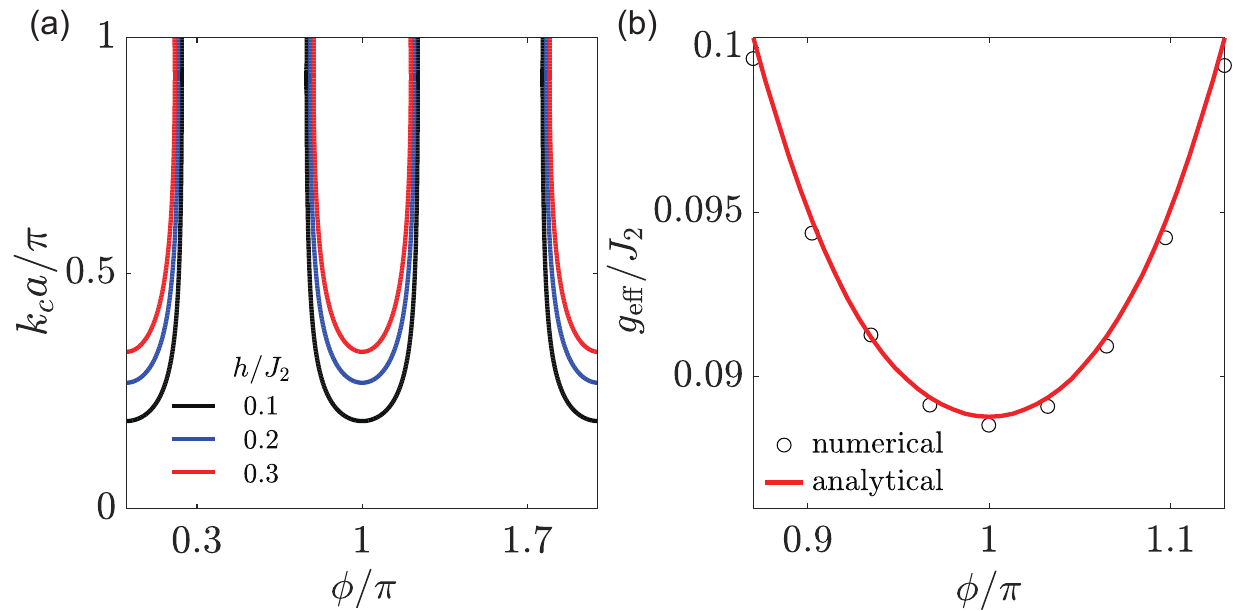}\\
 \caption{(a) The wave vector $k_c$ at crossing point as a function $\phi$ for different magnetic fields $h$. (b) The analytical (red curve) and numerical (black circles) results of the effective magnon-magnon coupling $g_{\rm{eff}}$ as a function of $\phi$. In the calculations, we set $J_1=0.4J_2$, $J_3=-1.25J_2$, and $K=0.05J_2$.}\label{fig5}
\end{figure}

In the above calculations, we focused on the case of magnons propagating along the $x$ direction. Notably, the anisotropic exchange interaction in altermagnets is expected to generate anisotropic magnon-magnon couplings. To illustrate this point, we express the magnon wavevector as $k(\cos\phi \hat{x}+\sin\phi \hat{y})$ with $\phi$ being the polar angle. We then derive the relation between the wavevector $k_c$ of the crossing point and the angle $\phi$,
\begin{equation}\label{Eq15}
\begin{aligned}
\cos(k_ca\cos\phi)-\cos(k_ca\sin\phi)&=\frac{h}{J_{1}-J_{2}} ,
\end{aligned}
\end{equation}which is a transcendental equation and it can only be solved numerically.

As plotted in Fig. \ref{fig5}(a), we observe that $k_c$ is symmetric about $\phi = 0$ and $\pi$ due to the mirror symmetry of the $y-z$ plane, while it takes the minimum when $\phi$ equals $0$ or $\pi$ because the difference in the group velocities of magnons with opposite chiralities reaches the maximum in such cases. When $\phi$ deviates from these two values, $k_c$ substantially increases. In addition, it is noted that Eq. \eqref{Eq15} has real solutions only when the magnon propagation angle $\phi$ falls into two windows $[-\phi_0,\phi_0]$ and $[\pi-\phi_0,\pi+\phi_0]$ with $\phi_0\approx\sqrt{4+2h/(J_1-J_2)}/\pi$, around $\phi = 0$ and $\pi$. Figure \ref{fig5}(a) also shows that the crossing point moves away from the origin as the magnetic field increases. A direct consequence of the $\phi$-dependence of $k_c$ is the anisotropy of the effective coupling $g_{\rm{eff}}$, as shown in Fig. \ref{fig5}(b). Numerical results compares very well with our analytical formula.

\textit{Discussion.---}Using the magnetic parameters of $\rm{RuO_2}$: $J_1= 0.8$ meV, $J_2 = 2$ meV, $J_3 = -3$ meV, and $K = 0.1$ meV \cite{Smejkal2}, one can estimate the effective magnon-magnon couping as $g_{\rm{eff}}=0.49$ meV for the spin density $N/V$ $\sim 10^{23}$ cm$^{-3}$ and $h=0.6$ meV. The resulting cooperativity, i.e., the ratio of the indirect coupling to the magnon dissipation rate, can reach $100$ for the Gilbert damping constant $\alpha=1.0 \times 10^{-3}$, indicating a strong coupling in the highly dispersive region. It is noted that the dipole-dipole interaction can break the degeneracy of magnons with opposite chiralities, while the resulting direct magnon-magnon coupling is negligibly small compared to the cavity-induced indirect coupling between exchange magnons \cite{Sheng2020,Tacchi2019}. Its contribution can be further dismissed when considering the long-distance coupling since the dipolar field generally decays as $1/d^3$ with $d$ the spatial distance of two magnets.

To summarize, we have predicted the cavity mediated long-range interaction between short-wavelength magnons in altermagnets. We showed that a perpendicular magnetic field can shift the magnon degeneracy from the origin to the exchange region, due to the unique chiral magnon splitting in altermagnets. The strong long-distance coupling between magnons with opposite handedness in cavities manifests as an anticrossing of the magnon-polariton spectrum in the highly dispersive region. We derived the analytical formula of the effective magnon-magnon coupling mediated by virtual photons and showed that it is highly anisotropic. Our results may open a new pathway for developing quantum magnonics and enrich the emerging research landscape of altermagnetism.

\begin{acknowledgments}
This work was funded by the National Key Research and Development Program under Contract No. 2022YFA1402802 and the National Natural Science Foundation of China under Grant No. 12074057.
\end{acknowledgments}

\end{document}